\documentclass[[journal,comsoc]{IEEEtran}
\usepackage{cite}
\usepackage{amsmath,amssymb,amsfonts}
\usepackage{algorithmicx,algpseudocode}
\usepackage{program}
\usepackage{algorithm}
\usepackage{graphicx}
\usepackage{textcomp}
\usepackage{xcolor}
\usepackage{multirow}
\ifCLASSINFOpdf
\else
\fi
\begin{document}
%
\title{Enhancing Security in VANETs with Efficient Sybil Attack Detection using Fog Computing\\}


\author{\IEEEauthorblockN{Anirudh Paranjothi,
Mohammed Atiquzzaman}
\IEEEauthorblockA{School of Computer Science, University of Oklahoma, Norman, Oklahoma, USA\\
e-mails: \{anirudh.paranjothi, atiq\}@ou.edu}}

\markboth{Journal of \LaTeX\ Class Files,~Vol.~14, No.~8, August~2015}%
{Shell \MakeLowercase{\textit{et al.}}: Bare Demo of IEEEtran.cls for IEEE Transactions on Magnetics Journals}
%



\IEEEtitleabstractindextext{%
\begin{abstract}
Vehicular ad hoc networks (VANETs) facilitate vehicles to broadcast beacon messages to ensure road safety. Rogue nodes in VANETs cause a Sybil attack to create an illusion of fake traffic congestion by broadcasting malicious information leading to catastrophic consequences, such as the collision of vehicles. Previous researchers used either cryptography, trust scores, or past vehicle data to detect rogue nodes, but they suffer from high processing delay, overhead, and false-positive rate (FPR). We propose a fog computing-based Sybil attack detection for VANETs (FSDV), which utilizes onboard units (OBUs) of all the vehicles in the region to create a dynamic fog for rogue nodes detection. We aim to reduce the data processing delays, overhead, and FPR in detecting rogue nodes causing Sybil attacks at high vehicle densities. The performance of our framework was carried out with simulations using OMNET++ and SUMO simulators. Results show that our framework ensures 43\% lower processing delays, 13\% lower overhead, and 35\% lower FPR at high vehicle densities compared to existing Sybil attack detection schemes.
\end{abstract}

}

\maketitle

\IEEEdisplaynontitleabstractindextext

%
\IEEEpeerreviewmaketitle

\section{Introduction}

Vehicular ad hoc networks (VANETs) have been emerging as a promising solution to intelligent transportation systems (ITS) with an aim to enhance road safety by reducing the number of road accidents and regulating traffic flow. In VANETs, the vehicles can communicate with each other using vehicle-to-vehicle (V2V) and vehicle-to-infrastructure (V2I) techniques, depends on dedicated short range communication (DSRC) [1]. The vehicles are equipped with on-board units (OBUs) for transmitting and receiving the messages between vehicles and between vehicles and the roadside units (RSUs). VANETs facilitate vehicles to broadcast beacon messages to disseminate network state or emergency information [2]. However, malicious vehicles, also known as rogue nodes may cause a Sybil attack to create an illusion of traffic congestion in order to reroute other honest nodes in the network to trap it or to cause catastrophic consequences, such as the collision of vehicles [3]. Therefore, efficient detection of the rogue nodes causing a Sybil attack is crucial in containing network damage and establishing a secure VANETs environment.

Previous authors used either cryptography, trust scores, or past vehicle data to detect rogue nodes. Baza et al. [4] presented an intrusion detection system (IDS) to detect rogue nodes based on the power of works (PoW) algorithm. Each vehicle starts the trajectory by transmitting the vehicle data using the public key and the signature using the private key to the RSUs. The RSUs verify the trajectory of the vehicles based on the historical vehicle data. If multiple trajectories for the same vehicle exists, the corresponding vehicle will be considered a rogue, and the information of the rogue node will be broadcasted to all vehicles in the region. The proposed scheme [4] suffers from a high delay and overhead in encrypting and decrypting the vehicle data. Zaidi et al. [5] proposed an IDS, where each vehicle utilizes its OBU to detect false data propagated by the rogue nodes based on the past vehicle data.  Ayaida et al. [6] proposed a trust model (TM) to detect false messages in VANETs based on the correctness of the data broadcasted in beacon messages. However, [5, 6] cannot be used in high vehicle dense regions, such as Manhattan and other downtown regions, due to high packet loss ratio (PLR) and false-positive rate (FPR).

To address the limitations of the existing Sybil attack detection schemes, we propose a fog computing-based Sybil attack detection for VANETs (FSDV). The FSDV framework employs the concept of guard nodes. The guard node is the vehicle that has more neighboring vehicles in its transmission range, dynamically creates a fog utilizing the OBUs of all vehicles in the region, and then the dynamic fog is used to analyze the received beacon messages from all vehicles to detect rogue nodes. FSDV exploits a traffic flow phenomena to generate a residual corresponding to the difference between the actual speed of the guard node and the received speed of the neighboring vehicles in a distributed way by using the beacon messages broadcasted over the network. A significant deviation of this residual from a dynamic threshold is considered as an indicator of the Sybil attack. The selection of the dynamic threshold directly impacts the Sybil attack detection probability. Thus, the threshold has to be chosen properly in order to well detect and to avoid false-positive detections. We adopt the fog computing technique to detect rogue nodes, as it offers low latency, low overhead, and high bandwidth compared to traditional communication techniques [7].

The \textit{novelty} of our proposed work lies in providing low data processing delays and low FPR at high vehicle densities. In addition, FSDV does not depend on any roadside infrastructures like RSUs or trust scores or past vehicle data in rogue nodes detection. The \textit{difference} between our framework and existing schemes [4-6] is, each vehicle uses its OBU or RSU to detect rogue nodes. RSUs are not uniquely deployed in all VANETs regions. Moreover, the overloaded RSUs yields high data processing delay and high FPR. OBUs of an individual vehicle is highly resource-constrained encounters a high delay in analyzing the data at high vehicle densities. Whereas, in the FSDV, the guard node combines OBUs of all vehicles in the region in creating the dynamic fog. Utilizing OBUs of all vehicles increases the computational power of dynamic fog resulting in low data processing delay and low FPR.

Our \textit{objective} is to reduce the latency in detecting rogue nodes and decrease FPR at high vehicle densities. We considered three existing rogue nodes detection schemes for comparison: PoW [4], IDS [5], and TM [6]. The performance of FSDV was carried out using OMNET++ and SUMO simulators with up to 4000 vehicles and 40\% rogue nodes. Our results lead to an exciting conclusion that our framework reduces the latency and FPR, and performs up to 32\% better than the existing Sybil attack detection schemes [4-6].

The \textit{contributions} of the paper are: 1) We proposed a framework, FSDV, that uses fog computing technique to detect Sybil attacks in VANETs with low delay and low FPR. 2) We introduced the guard node in FSDV, which uses an OBU-based fog computing technique to compare speed values received in the beacon messages from all vehicles in the region. 3) We performed an extensive simulation by creating a dynamic fog layer under varying vehicular and network conditions to determine false information broadcasted by rogue nodes.

The rest of the paper is structured as follows: related work is discussed in Section II. The working principle of our proposed work to detect rogue is illustrated in Section III. The mathematical model analysis is discussed in Section IV. Section V presents the performance evaluation of the proposed framework. The simulation results are discussed in Section VI before concluding the paper in Section VII.

\section{Related Work}

This section presents an overview of the most recent existing schemes that detect rogue nodes in VANETs. Arshad et al. [8] and Ahamad et al. [9] proposed a trust-based scheme to detect rogue nodes in VANETs.  Based on the correctness of the data in beacon messages, positive or negative trust values are assigned. Positive and negative trust values represent the normal and abnormal behavior of the vehicles, respectively. When the calculated trust of any vehicle reaches a predefined threshold limit is known as a rogue node and then the information is broadcasted to all the vehicles in the region. 

Iwendi et al. [10] proposed a spider money technique, which utilizes the RSUs in the region to authenticate the private key associated with beacon messages to identify malicious data broadcasted by the rogue nodes. Once the rogue nodes are identified, the RSUs send rogue nodes information to the department of motor vehicles (DMV) to withdraw the vehicle's permit to reduce the potential damage to the network. However, the frameworks [8-10] encounter a low true-positive rate (TPR) and high FPR in detecting rogue nodes. 

Al-Otaibi et al. [11] presented a cryptography-based fog computing scheme to detect rogue nodes in VANETs. The RSUs are considered as fog nodes for rogue nodes detection. However, this approach [11] encounters a high processing delay and overhead in detecting rogue nodes when the RSUs are overloaded in the region. Yang et al. [12] proposed a machine-learning algorithm to classify the historical vehicle data received from all the vehicles in the region. The result of the classification, i.e., whether the data received from the vehicle is valid or not, are then combined to detect rogue nodes broadcasted malicious information. However, this approach [12] encounters high delay and high overhead. 

To overcome the limitations of the existing Sybil attack detection schemes [4-6, 8-12], we propose the FSDV framework, which utilizes only vehicle speed values in beacon messages and does not depend on either trust score, cryptography, or past vehicle data in rogue nodes detection.

\section{Proposed FSDV Framework}

\begin{figure}[tbp]
\centering
\includegraphics[width=250pt]{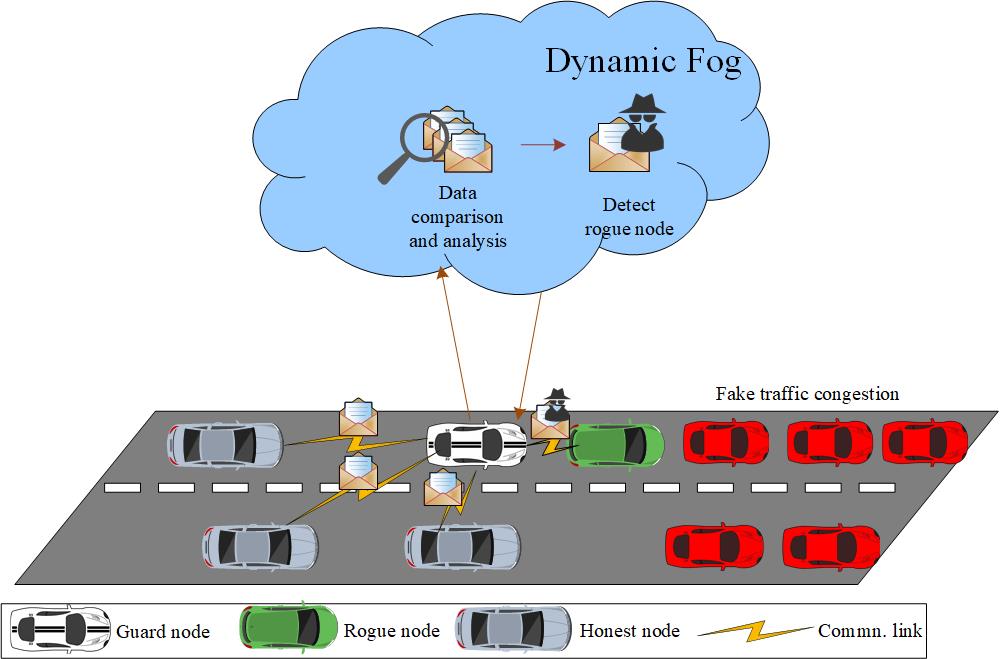}
\caption{Execution scenario of FSDV in the presence of a rogue node using fog computing technique.\label{fig1}}
\end{figure}

In this section, we discuss the working principle of our FSDV framework. The FSDV framework engages fog computing in detecting rogue nodes involved in Sybil attacks. The rogue nodes broadcast multiple messages with different vehicle id to all vehicles in the region, thereby giving a false impression of fake traffic congestion ahead by lowering the speed values in the beacon messages. One such scenario of Sybil attacks is illustrated in Fig. 1.

In this paper, we employ the concept of guard node to detect Sybil attacks in VANETs. The vehicle which has more neighboring vehicles in its communication range will act as a guard node. Guard node dynamically creates a fog utilizing OBUs of all vehicles to detect rogue nodes broadcasting lower speed values in the region. Once the fog layer is deployed, the guard node starts adding the neighboring vehicles to the list of neighbors, as met for the first time. Otherwise, it updates the timestamp of the neighboring vehicles. The neighbors list updated continuously when the beacon messages are broadcasted in the region (i.e., for every 100ms). Then, the detection of rogue nodes is achieved by comparing and analyzing the speed values broadcasted in beacon messages. If the speed difference is greater than the dynamic threshold, a Sybil attack is detected. This is due to all the vehicles in the region broadcast similar speed values as they are in similar traffic conditions and dependent on other vehicles under all circumstances. Thus, if there is a significant difference in vehicle speed, the guard node considers the vehicles as rogue nodes. Otherwise, the vehicles are highlighted as honest nodes. 

The guard node broadcasts the information of rogue nodes to all vehicles in the region to ignore the beacon messages received further from the rogue nodes to contain the network damage. The adoption of fog computing increases the computation power of the guard node as the OBUs of all vehicles are utilized in creating a dynamic fog results in lower data processing delay compared to [4-6] schemes.

\subsection {Selection of Guard Node}

Rogue nodes are the vehicles broadcasting low-speed values or sudden decrease in speed to change the normal behavior of the vehicles for own benefits. In the FSDV framework, the guard node analyzes the speed values in the beacon messages received from all vehicles in the region to detect rogue nodes. The following two assumptions are made in the selection of the guard node: First, we assume the center vehicle in the region will act as a guard node ($G_{n}$), as the vehicle in the center has more number of the neighboring vehicles in its communication range compared to the front and tail-end vehicles. Second, we assume that the total number of vehicles (\textit N) in the region at any given time is at least two. The guard node needs at least two vehicles in the region to compare and analyze the beacon messages to detect rogue nodes causing a Sybil attack.

Initially, we take the mean of position vectors of all vehicles (i.e., $Pt_1$, $Pt_2$, ...., $Pt_N$)  to find a unique center point $\zeta$. 

\begin{equation} \label{eq11}
 \begin{split}
 \zeta = \frac{1}{N} \sum_{i=1}^{N} Pt_i
 \end{split}
 \end{equation}

We calculate Euclidean distance between $\zeta$ and the position vector of each vehicle and then determine the point that has the minimum distance from the $\zeta$. Finally, the vehicle located at this point will be selected as the guard node, $G_{n}$. 

\begin{equation} \label{eq12}
 \begin{split}
 G_{n} = \arg\min_{Pt_i \in Y} \|\zeta - Pt_i\|
 \end{split}
 \end{equation} 
 
  Where, $Y$ = \{$Pt_1$, $Pt_2$....,$Pt_N$\}.

\subsection {Density and Speed of Vehicles}

In the FSDV framework, Greenshield's traffic flow model is used to calculate the speed of the guard node and the density of the vehicles in the region. Greenshield traffic model is considered to be a fairly accurate and simple model for real-world traffic flows works under the assumption of density ($\rho$), and the speed of the vehicles (\textit S) is negatively correlated [5]. The density can be calculated as:

\begin{equation} \label{eq3}
 \begin{split}
 \rho = B_{msg} \cdot N
 \end{split}
 \end{equation} 
 
 Where, $B_{msg}$ is the beacon message received from one vehicle id and $N$ is the total number of vehicles in the region. As the speed and density of the vehicles are negatively correlated, the density increases when the speed of the vehicles decreases in the region. The speed of the guard node ($S_G$) can be defined as:

 \begin{equation} \label{eq5}
 \begin{split}
 S_G = S_{max} -  \frac {\rho}{\rho_{max}} S_{max}
 \end{split}
 \end{equation}

 Where $S_{max}$ is the speed of the vehicle when density is zero and ${\rho_{max}}$ is the maximum density, also point at which speed of the vehicles becomes zero. The calculated speed value of the guard node ($S_G$) is compared with the received speed values of the neighboring vehicles ($S_{rcvd}$). If the difference in speed values is greater than the dynamic threshold ($S_{th}$), i.e., $S_G$ -- $S_{rcvd}$ $>$ $S_{th}$, a Sybil attack is detected, and the vehicle broadcasted corresponding speed values is considered as a rogue node. Thus, the guard vehicle broadcasts the information of rogue nodes to all the vehicles in the region.

\subsection {FSDV Algorithm}
 
\begin{algorithm}
\caption{ FSDV Sybil Attack Detection algorithm }\label{alg1}
\textbf{Input:} $G_{n}$ receives $B_{msg}$ from all vehicles in the region
\newline
\textbf{Output:} $G_{n}$ broadcasts information of rogue nodes

\begin{algorithmic} [1]
\If {($N$ $\geq$ 2)} 
\State Calculate  $\zeta$ and Euclidean distance 
\State Assign  $G_{n}$ 
\Else \, GoTo 22
\EndIf
\State $G_{n}$ dynamically creates a fog 
\State $G_{n}$ receives $B_{msg}$ from all vehicles in the region
\For {each  $B_{msg}$  received}
	\If {($B_{msg}$.${Sender} \not\in {Neighbors_{list}}$)}
		\State ${Neighbors_{list}}$.add {($Sender$)}
	\EndIf 
	\State ${Neighbors_{list}}${[$Sender$]}.{$T_{stamp}$} =  $B_{msg}$.{$T_{stamp}$} 
	\State Calculate $S_G$
	\If {($S_G$ -- $S_{rcvd}$ $<$ $S_{th}$)}
	\State Declare the vehicle as honest node
	\Else
	\State Declare the vehcile as rogue node
	\State Store the rogue node id
	\EndIf
\EndFor 
\State $G_{n}$ broadcasts rogue nodes information through $B_{msg}$ 
\State Terminate the Sybil attack detection algorithm
\end{algorithmic}
\end{algorithm}
 
 \section{Mathematical Model Analysis}

\begin{figure}[tbp]
\begin{center}
\includegraphics[width=220pt]{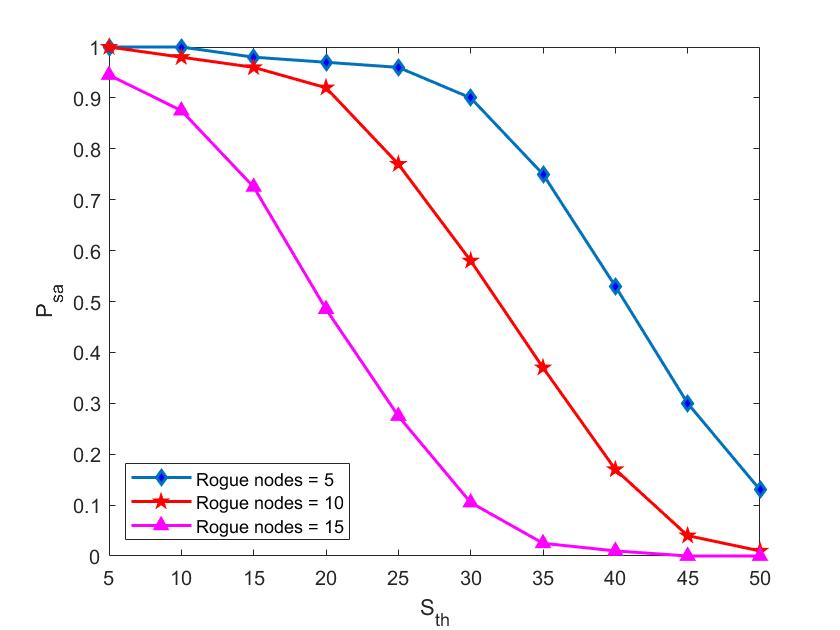}
\caption{Effect of dynamic speed threshold in the Sybil attack detection probability.}
\end{center}
\end{figure}
 
To validate the performance of the proposed approach, discussed in Section III, we investigate in this section the theoretical expression for the delay and the probability of detecting the rogue nodes correctly in the Sybil attack.

\subsection {Analysis of Delay in FSDV}

As our objective is to reduce the latency significantly in detecting rogue nodes compared to [4-6] schemes, we have analyzed different types of delays, such as communication delay, queuing delay, and data processing delay associated with the FSDV framework, illustrated in Section III. The communication delay, $D_c$ associated with the guard node in transmitting the received beacon messages ($B_{msg}$) from all vehicles in the region to the fog layer is given by: 


 \begin{equation} \label{eq5}
 \begin{split}
   D_c = \frac{x}{r} = \frac{x}{b \log_2\left(1+ \frac{tc^2}{\sigma^2} \right)}
 \end{split}
 \end{equation} 
 
Where, $x$ is the number of bits in the beacon messages, $t$ is the transmission power of fog devices at the fog layer, $c$ is the channel coefficient between devices the guard node and the fog layer, ${\sigma^2}$ is the power of white Gaussian noise, $r$ is the transmission rate calculated from Shannon channel capacity theorem, and $b$ is the bandwidth of the link established between the guard node and the fog layer. 

Devices associated with the fog layer use the M/M/1 queuing model for preparing the beacon messages to be computed, where the arrival and service time are represented as $\frac{1}{\lambda_1}$ and $\frac{1}{\mu_1}$, respectively. The queuing delay ($D_q$) associated with a fog layer of our FSDV framework is given by:

 \begin{equation} \label{eq5}
 \begin{split}
   D_q = 1 - \frac {1} {\lambda_1} +\frac {{\lambda_1}^2} {\mu_{1}^2(\mu_{1} - \lambda_1)}
 \end{split}
 \end{equation} 
 
The data processing delay ($D_p$) indicates the required time to process all received beacon messages in the fog layer to detect rogue nodes is given by:
 
 \begin{equation} \label{eq5}
 \begin{split}
   D_p = \frac {{TC(x)}f}{c_{fl}}
 \end{split}
 \end{equation} 
 

Where $c_{fl}$ is the computation capability of the fog layer, $f$ is the number of CPU cycles required for computing one bit of data at the fog layer, and $TC(x)$ is the time complexity for $x$ bits of data in the beacon messages, depends on the algorithm discussed in Section III. The total delay ($D_t$) of our FSDV framework in the fog layer is given by:

 \begin{equation} \label{eq5}
 \begin{split}
   D_t = D_c + D_q + D_p
 \end{split}
 \end{equation}

\subsection{Rogue Nodes Verification}

Assume the beacon messages from all other vehicles in the region are successfully received by the guard vehicle. The probability of comparing the speed values in the beacon messages to detect the rogue nodes correctly is given by:

 \begin{equation} \label{eq5}
 \begin{split}
   X(P_1P + P_2P)
 \end{split}
 \end{equation} 
 
 The parameter $P$ is the probability that beacon messages positively reach the guard vehicle from all other vehicles in the region. $P_1$ represents the probability of predicting the honest node correctly if the difference between the calculated speed of the guard node and the honest node is lower than the dynamic threshold (i.e., $S_G$ -- $S_{rcvd}$ $<$ $S_{th}$ ). $P_2$ represents the probability of predicting the rogue node correctly if the difference between the calculated speed of the guard node and the rogue node is greater than the dynamic threshold (i.e., $S_G$ -- $S_{rcvd}$ $>$ $S_{th}$). $X$ represents the probability that the guard node successfully creates a dynamic fog layer to compare and analyze the speed values in the beacon message to detect the rogue nodes in the region. The probability of incorrectly predicting the rogue nodes, $P_i$ is given by:
 
  \begin{equation} \label{eq5}
 \begin{split}
   P_i &= 1 - [XP_1P + XP_2P] \\
   	  & = 1 - [XP_1 + XP_2]P  \\
   	  & \approx 1-XP_2P
 \end{split}
 \end{equation} 
 
\section{Performance Evaluation}

This section evaluates and analyzes the performance of the FSDV framework discussed in Section III.

\subsection {Evaluation of Speed Threshold in FSDV}

As discussed in Section III, a significant deviation of the speed values from a dynamic threshold ($S_{th}$) is considered as a Sybil attack. The threshold value is more dynamic as it depends on the speed of the vehicles in the region. Thus, when the speed of the vehicle decreases, the dynamic threshold value also decreases in the region. For example, the dynamic threshold value in the dense vehicle region, such as the downtown region is lower, compared to the dynamic threshold value in the fluid traffic region. This is due to the speed of the vehicles in a high dense vehicle region is lower compared to the fluid traffic region. The Fig. 2 shows how the selected speed threshold $S_{th}$ impacts the detection probability $P_{sa}$. 

From Fig. 2, we can infer that the probability of Sybil attack detection is high when the speed threshold ($S_{th}$) is low. This is due to we were able to identify all the rogue nodes in the region. However, the low-speed threshold may hide false positive detection. When the speed threshold ($S_{th}$) increases, the detection probability decreases since we can miss some rogue nodes due to the high-speed threshold. Therefore, to detect all the rogue nodes in the region and to reduce FPR, the speed threshold has to be chosen dynamically based on the speed of the vehicles in the region.

\begin{table}
\centering
\caption{Parameters used in Simulation of the FSDV Framework\label{tab1}}
\begin{tabular} [htbp] {|c|c|}

\hline

\textbf{Parameters} & \textbf{Values} \\    \hline \hline

Number of vehicles & 500-4000 \\    \hline
Road length & 5 Miles \\    \hline
Number of lanes & 2 \\    \hline
Vehicle speed &  30-65 Miles/hr\\    \hline
Transmission range & 500 m\\    \hline
Beacon message size & 256 bytes \\    \hline
Protocol & IEEE802.11p \\    \hline

\end{tabular}
\end{table}

\subsection {Simulation Setup}

The simulations are carried out using OMNET++ and SUMO simulators. SUMO is an open-source traffic-events simulator, provides a trace of vehicle movements, such as vehicle speed, position, acceleration, etc. at the end of every simulation for a map imported from OpenStreetMap. OMNET++ is a discrete event simulator, used to measure the performance of the network using the node deployment model, node mobility model, etc. To perform the simulation, we imported the map of the city of Norman, Oklahoma into the SUMO simulator. The output of the SUMO simulator, i.e., the trace of vehicles, is given as input to the OMNET++ simulator for rogue nodes detection. To assess the scalability and behavior of the FSDV framework, we increase the number of vehicles up to 4000 and the presence of rogue nodes in the network up to 40\%. Table I summarizes the most commonly used parameters used in the simulation.

\subsection {Performance Metrics}

The simulations were performed based on the equations formulated in Section III and IV. We considered the following metrics to evaluate the performance of our FSDV framework and to compare our results with PoW, IDS, and TM schemes:

\begin{itemize}

\item Data processing time: The time needed by the guard node to analyze the beacon messages received in the region.

\item PLR: The ratio of the number of lost packets to the total number of packets sent across a communication channel.

\item Average throughput: Average rate of successfully broadcasted beacon messages across a communication channel. 

\item Overhead: The additional information exchanged between the vehicles to detect rogue nodes in the region.

\item True positive rate: The percentage of rogue nodes is accurately detected and classified as rogue nodes.

 \begin{equation} \label{eq13}
 \begin{split}
\text {TPR} = \frac{\text {No. of rogue nodes detected correctly}}{\text{Total no. of rogue nodes}}
 \end{split}
 \end{equation}

\item False positive rate: The percentage of honest nodes is incorrectly detected and classified as rogue nodes. 

 \begin{equation} \label{eq13}
 \begin{split}
\text {FPR} = \frac{\text{No. of honest nodes detected incorrectly}}{\text{Total no. of honest nodes}}
 \end{split}
 \end{equation}

\end{itemize}

 \begin{figure*}[tbp]
\centering
\includegraphics[width=475pt]{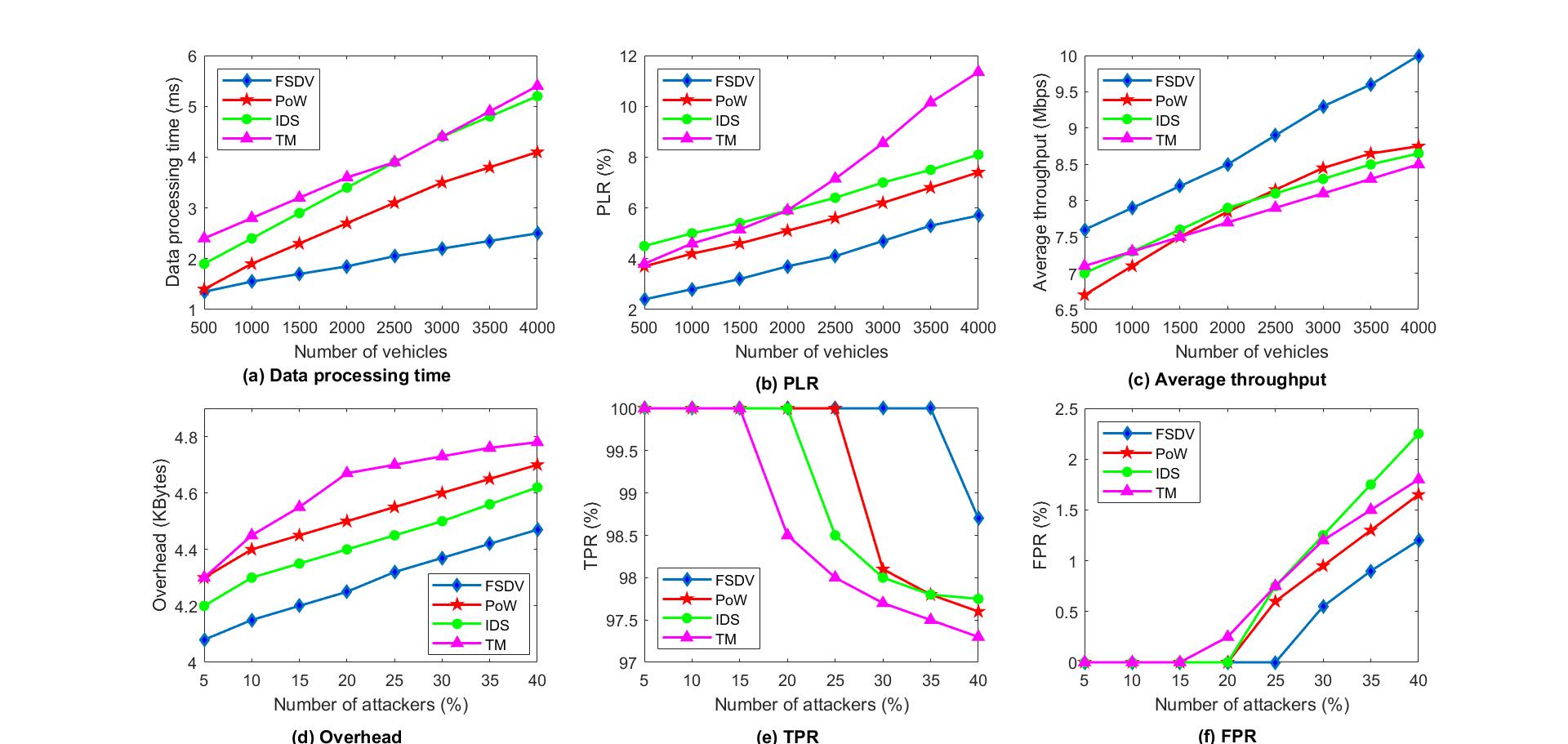}
\caption{Comparison of the FSDV framework with PoW, IDS, and TM schemes: (a) data processing time, (b) PLR, (c) average throughput, (d) overhead, (e) TPR, (f) FPR. \label{fig3}}
\end{figure*}

\section{Results}
During the simulation, the FSDV framework is analyzed based on the equations formulated in Section III and Section IV, and the performance metrics illustrated in Section V. The results are presented below:

\textit{1) Data processing time:} The time taken by the guard node to process speed values in the received beacon messages (Eq. 7). When the number of vehicles increases from 500 to 4000, the data processing time increases as the guard node needs to process a large number of beacon messages received from all the vehicles in the region. However, as the OBUs of all vehicles are utilized in creating the dynamic fog layer, the computation power of the guard node increases when the number of vehicles increases in the region results in a 43\% lower processing delay compared to [4-6] schemes. In the 4000 vehicles simulation, the data processing time is 39\%, 51\%, and 54\% lower than PoW, IDS, and TM schemes, respectively, as shown in Fig. 3a. The results show that our FSDV framework is efficient and can handle high vehicle densities.

\textit{2) PLR:}  PLR of the FSDV framework increases when the number of vehicles increases from 500 to 4000 due to the high mobility of the vehicles resulting in a collision of some packets. However, the PLR of our framework is lower compared to [4-6] at all vehicle densities. The PLR is calculated against the number of vehicles. In 4000 vehicles simulation, PLR is 22\%, 29\%, and 38\% lower than the PoW, IDS, and TM schemes, respectively, as shown in Fig. 3b.

\textit{3) Average throughput:} The average throughput is calculated against the number of vehicles and increases when the number of vehicles increases, as shown in Fig. 3c. Due to a large number of messages are successfully broadcasted to all vehicles region, the average throughput of the FSDV framework is higher compared to [4-6] schemes at all vehicle densities.  In the 4000 vehicle simulation, average throughput is 14\%, 15\%, and 19\% higher than the PoW, IDS, and TM, respectively.

\textit{4) Overhead:} The overhead of our framework is calculated against the number of rogue nodes (Fig. 3d). Finding the dynamic threshold increases the overhead of our framework (Section V). However, the adoption of the fog computing technique yields lower network overhead compared to traditional communication techniques. Thus, the overhead of the FSDV framework is 13\% lower compared to [4-6] schemes. For a network with 40\% rogue nodes, the overhead is 7\%, 6\%, and 10\% lower than the PoW, IDS, and TM schemes, respectively.

\textit{5) TPR:}   It is difficult to detect the rogue nodes broadcasting false information if the speed varies gradually in the received beacon messages. However, to generate either fake traffic congestion or accident, the target rogue node decreases the speed values quickly. Therefore, our FSDV framework detects rogue nodes correctly (i.e.,100\%) up to 30\% rogue nodes in the region and decreases slightly to 98.7\% when the number of rogue nodes increased to 40\%, as shown in Fig. 3e. Moreover, the TPR of the FSDV framework is higher compared to PoW, IDS, and TM schemes at all vehicle densities.

\textit{6) FPR:} The increase in FPR increases rogue nodes in the region as well as deteriorates the performance of the rogue node detection schemes (Section V). In the FSDV framework, the rogue nodes detection relies only on the speed values in beacon messages broadcasted by all vehicles in the region without taking into any trust scores or past vehicle data into consideration resulting in 35\% lower FPR compared to [4-6] schemes. For a network with 40\% rogue nodes, the FPR is 31\%, 51\%, and 38\% lower than PoW, IDS, and TM schemes, respectively, as shown in Fig. 3f.

\section{Conclusions and Future Work}

We studied the challenges in detecting Sybil attacks, such as high processing delay, high network overhead, low TPR, and high FPR, notably when the number of rogue nodes increases in the region. To address these limitations and to provide an efficient scheme to detect false messages broadcasted over the network, we proposed an OBU-based fog computing technique called FSDV to detect rogue nodes increases by up to 40\% in the region. The simulations were performed to evaluate the performance of the FSDV framework using OMNET++ and SUMO simulators. Results showed that the FSDV framework is scalable, efficient, robust, and performs up to 32\% better than [4-6] techniques. Moreover, the FSDV framework ensures a 43\% lower processing delay, 35\% lower FPR, and 13\% lower overhead at high vehicle densities compared to existing Sybil attack detection schemes [4-6]. 

Our framework does not depend on any roadside infrastructures like RSUs or trust scores or past vehicle data in rogue nodes detection, which is a major advantage compared to existing schemes [4-6, 8-12]. In the future, we are planning to extend this work to detect various attacks, such as tunneling attacks, impersonation attacks, etc. This can be done by simulating the environment of the security attacks and then detecting the malicious information broadcasted using rogue nodes detection techniques.




\end{document}